# *Distinct Structural Dynamics of the Semiquinone State Define a Signalling Pathway in Avian Cryptochrome*

*Monika Kish* [a,c,†], *Suchitra Pradhan* [a,b,d,†], *Jessica L. Ramsay* [a,b,e,†], *Paloma Munguía Salazar* [a], *Jonathan Phillips* [a,c,f,*], *and Daniel R. Kattnig* [a,b,*]

†: These authors contributed equally (alphabetical order).

*: jj.phillips@exeter.ac.uk, d.r.kattnig@exeter.ac.uk

[a] *Living Systems Institute, University of Exeter, Stocker Road, Exeter, Devon, EX4 4QD, United Kingdom.*

[b] *Department of Physics, University of Exeter, Stocker Road, Exeter, Devon, EX4 4QL, United Kingdom.*

[c] *Department of Biosciences, University of Exeter, Stocker Road, Exeter, Devon, EX4 4QD, United Kingdom.*

[d] *Current affiliation: Bristol Veterinary School, University of Bristol, Bristol, United Kingdom.*

[e] *Current affiliation: Division of Genetics and Genome Biology, School of Biological and Biomedical Sciences, University of Leicester, Leicester, LE1 7RH, United Kingdom.*

[f] *Alan Turing Institute, British Library, London, NW1 2DB, United Kingdom.*

**Abstract**

The light-dependent magnetic compass of night-migratory songbirds is widely hypothesized to rely on the radical pair mechanism within retinal cryptochrome. However, bridging the mechanistic gap between microsecond quantum spin dynamics and the long-lived, global protein conformational changes required for cellular signalling remains a formidable challenge. Here, we apply redox state-resolved hydrogen/deuterium-exchange mass spectrometry (HDX-MS) to map the conformational landscape of European robin cryptochrome 4a (ErCry4a) across its photocycle. We reveal that photochemical reduction drives robust, allosteric structural transitions across key functional nodes, including the phosphate-binding loop (PBL), protrusion loop (PL), FAD-proximal helix α17, and the C-terminal α22/α23 network. Crucially, we isolate the structural fingerprint of the transient semiquinone, the presumed signalling species. Rather than acting as a linear structural stepping-stone, the semiquinone exhibits a distinct, non-monotonic conformational signature characterized by a transient destabilization of the PBL and PL, contrasting sharply with the global rigidification observed in the fully reduced state. These findings establish the semiquinone as a structurally unique and functionally competent biological entity. Our results provide direct biophysical evidence for a dedicated, high-fidelity structural signalling cascade, detailing how localized quantum-level photochemistry is translated into the precise conformational dynamics required for animal navigation.

## Introduction

Magnetoreception enables diverse animal species to utilize the geomagnetic field for navigation and orientation [1]. While widespread across the animal kingdom, this sensory modality has been most extensively characterized in night-migratory songbirds, which rely critically on a light-dependent magnetic compass [2, 3]. Mounting behavioural and biophysical evidence, alongside rigorous theoretical modelling, points to the radical pair mechanism as the physical basis of this compass. This model posits that magnetic sensing is underpinned by the quantum-coherent spin dynamics of photoinduced radical pairs within cryptochrome flavoproteins localized in the avian retina [4]. However, directly linking the mechanism to *in vivo* physiological signalling remains a formidable challenge, with unambiguous physico-chemical insight limited to a handful of *in vitro* studies.

*In vitro* studies have successfully identified magnetically sensitive, light-induced radical pairs across several members of the cryptochrome/photolyase family. These include AtCRY1 from *Arabidopsis thaliana* [5, 6], DmCry from *Drosophila melanogaster* [7], and ErCry4a [8], a cryptochrome 4 variant from the European robin (*Erithacus rubecula*), the species in which compass magnetoreception was originally discovered [9]. In these studies, the primary magnetosensitive species is the $[\mathrm{FAD}^{\bullet-} \cdots \mathrm{TrpH}^{\bullet+}]$ radical pair, comprising a semi-reduced, non-covalently bound flavin adenine dinucleotide (FAD) chromophore and a surface-exposed oxidized tryptophan (Trp) radical, generated via sequential electron transfer through a conserved chain of tryptophan residues (cf. Fig. 1). While this canonical pair exhibits robust sensitivity to moderately high magnetic field strengths (~ 10 mT) *in vitro*, its physiological relevance is still debated [10]. A prominently discussed alternative is the flavin semiquinone–superoxide radical pair, $[\mathrm{FADH}^{\bullet} \cdots \mathrm{O_2}^{\bullet-}]$, which could putatively form during dark-state re-oxidation [11, 12]. However, while recent theoretical frameworks suggest this superoxide-containing pair might (unexpectedly) possess low-field sensitivity [13], direct *in vitro* experimental validation remains absent.

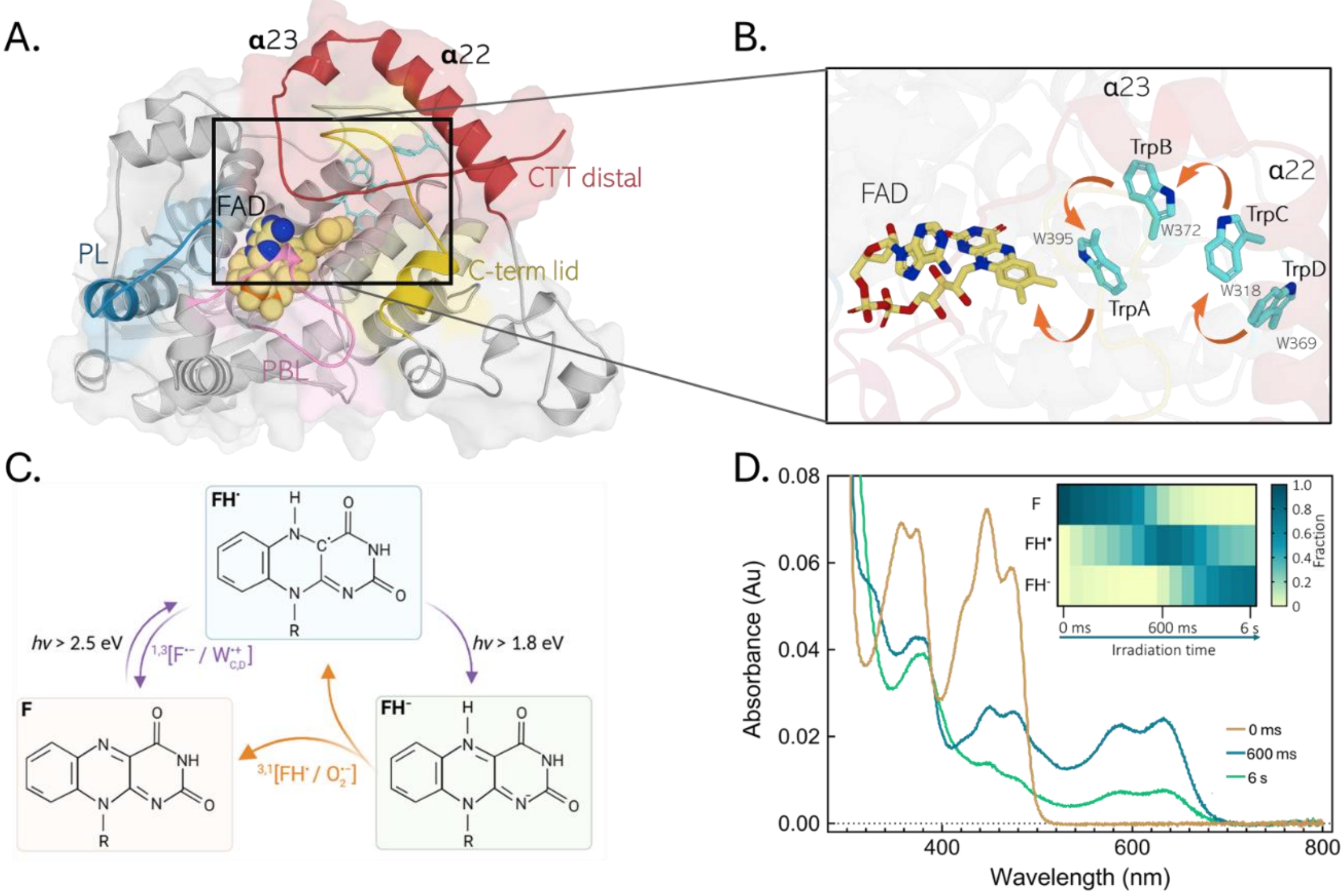

**Figure 1. Structure, electron transfer pathways, and redox cycle of European robin cryptochrome 4a – the semiquinone state of cryptochrome is transiently enriched by millisecond blue light stimulation.** (A) Structural model of ErCry4a based on ClCry4a (PDB: 6PU0) with polypeptide cartoon coloured to highlight functional regions: phosphate binding loop (PBL) – pink; protrusion loop (PL) - cyan; C-terminal lid – yellow; C-terminal tail (CTT) – red; flavin adenine dinucleotide (FAD) – atomic colours. (B) Electron transfer chain via tryptophan residues supports photoreduction of FAD. (C) Photoreduction / re-oxidation cycle of FAD. Redox state changes involve radical pair intermediates, the recombination of which is established ($[F^{\bullet-}/W^{\bullet+}_{C,D}]$) or hypothesised ($FH^{\bullet}/O_2^{\bullet-}$) to be magnetic field sensitive. (D) Representative UV/Vis spectra for time resolved photoreduction experiments at point of maximum enrichment for the fully oxidised (F), semiquinone ($F^{\bullet}$) and fully reduced ($FH^{-}$) states. Inset is calculated fraction of each state to total spectral signal during time resolved photoreduction experiments, showing maximum of $F^{\bullet}$ enrichment at ~600 ms. Spectra were recorded in 50 mM HEPES buffer (pH 8.0) containing 2 mM dithiothreitol (DTT) and 10% glycerol at room temperature, with irradiation at 450 nm.

Irrespective of the specific *in vivo* radical partner, the unifying feature of magnetosensitive cryptochromes is the transition of the non-covalently bound FAD chromophore between its distinct redox states [12, 14]: the fully oxidized flavin (FAD), the one-electron-reduced semiquinones ($FAD^{\bullet-}$ or $FADH^{\bullet}$), and the fully reduced hydroquinone ($FADH^{-}$) (Fig. 1 (C)). These redox transitions proceed via transient radical pair intermediates, whose chemical fate is governed by the quantum-coherent interconversion between singlet and triplet spin states [4]. External magnetic fields perturb these spin dynamics, effectively altering the quantum yields of radical pair recombination relative to the forward progression into longer-lived product states [4, 6].

To transduce this localized magnetic information into a biochemical cellular response, the protein must populate a discrete signalling state capable of biochemical interaction with downstream binding partners [15, 16]. This necessitates a physical coupling between the FAD redox state and defined conformational rearrangements within the cryptochrome architecture [12, 17, 18]. Implicit in this transduction mechanism is a profound challenge of bridging timescales: the primary magnetosensitive spin dynamics, constrained by radical pair decoherence to the microsecond regime [19, 20], must be translated into cellular signalling networks that operate over milliseconds to hours. Consequently, a biologically viable magnetoreceptor must structurally stabilize its signalling-active state - presumably the semiquinone state - for at least the millisecond duration required to initiate downstream signal propagation.

Despite these mechanistic imperatives, the specific pathway by which photochemical events are transduced into structural rearrangements remains poorly defined for cryptochromes from migratory species. While avian Cry1a has historically been proposed as a candidate magnetoreceptor [21], its weak *in vitro* FAD binding affinity [22] and lack of light-dependent activation [14, 23-25] dispute its role as a primary photosensor. Consequently, Cry4a, the only avian cryptochrome for which magnetic field effects on photoreduction have been explicitly characterized to date, stands as the prime, biologically viable model for this compass mechanism. Yet, our current structural framework for ErCry4a relies largely on functional extrapolation from the well-characterized activation of plant cryptochromes [26-28] and, to a lesser extent, those of *D. melanogaster* [29-31]. Initial investigations into pigeon and chicken cryptochrome 4 indicate that light excitation triggers a reversible conformational shift, primarily characterized by a decreased solvent exposure of the C-terminal tail (CTT) [12, 14, 17]. For pigeon ClCry4 in particular, limited proteolysis experiments suggest a light-induced disorder-to-order transition within the phosphate-binding loop (PBL), potentially coupled to CTT dynamics [12]. Structural model of ErCry4a based on ClCry4a (PDB: 6PU0 [12]) with polypeptide cartoon coloured to highlight functional regions can be seen in Figure 1 (A). However, these foundational observations carry significant caveats. The limited spatial resolution of proteolysis assays, combined with steady-state illumination protocols, precludes the assignment of these structural transitions to a specific, functionally relevant FAD redox state. Furthermore, it is difficult to definitively uncouple these observed effects from potential photodegradation artifacts, and the evolutionary conservation of these structural dynamics in the cryptochromes of strictly night-migratory species remains unconfirmed. Crucially, emerging biochemical data suggest that the signalling interface of ErCry4a extends well beyond the C-terminus. Of six recently identified interaction partners for ErCry4a, only half coordinate with the CTT [15]. Most notably, the CTT is entirely dispensable for ErCry4a's interaction with cone-specific G-proteins [16]. Taken together, these findings highlight a critical knowledge gap: the global structural dynamics that define the actively signalling, magnetoreceptive state of ErCry4a are still largely unresolved.

To resolve this structural ambiguity, we employed hydrogen/deuterium-exchange mass spectrometry (HDX-MS) to map the conformational landscape of ErCry4a as a function of its specific FAD redox state. Unlike low-resolution structural probes, HDX-MS offers a highly sensitive, label-free approach to quantify backbone amide solvent accessibility, providing a direct readout of local structural dynamics and transient unfolding events [32, 33]. Crucially, recent methodological advancements now permit the interrogation of large, dynamic protein systems with millisecond kinetic resolution [34-36] and near-single-residue structural resolution [37-41], making it uniquely suited to capture the transient conformational states of magnetosensitive cryptochromes. Despite this potential, the application of HDX-MS to the cryptochrome family remains remarkably limited and, to date, entirely restricted to non-vertebrate species. The three existing studies—investigating *Phaeodactylum tricornutum* CryP [42], *Chlamydomonas reinhardtii* CraCry [43], and *Drosophila melanogaster* DmCry [31] - have revealed light- and redox-dependent structural mobilizations primarily clustered near the FAD-binding pocket (notably helices α16/α17) and the structural elements connecting to the C-terminal region (such as helices α22–α23). While species-specific mechanistic variations undoubtedly exist, the recurring involvement of these FAD-proximal and C-terminal-linked regions establishes them as highly conserved nodes of structural communication. By leveraging state-resolved HDX-MS on a *bona fide* vertebrate magnetoreceptor, we aim to definitively trace this signalling pathway from the initial neutral semiquinone formation to the global protein response. More broadly, we test the conjecture that quantum effects can be transduced into a structural change in the protein.

## Results

### *Protein expression, characterisation, and photocycle*

To investigate the redox-induced structural dynamics of ErCry4a, we recombinantly expressed and purified the wild-type protein from *E. coli* using an optimized protocol based on prior methods [8]. The purified fraction exhibited the characteristic deep yellow colour indicative of bound, fully oxidized FAD, which was confirmed by optical spectroscopy revealing the canonical vibronic structure with an absorption maximum at 450 nm (Fig. S1). Intact mass spectrometry confirmed sample integrity, displaying a major species at approximately 62,940 Da, in excellent agreement with the theoretical mass of the full-length, His-tagged construct (Fig. S2). A secondary peak at 62,810 Da corresponds to a mass shift of -130 Da, consistent with the standard post-translational cleavage of the N-terminal methionine; a minor third peak at 63,120 Da likely reflects an additional, low-abundance post-translational modification. For subsequent HDX-MS analysis, enzymatic digestion yielded 266 identifiable peptides, achieving an exceptional sequence coverage of 99.1% with an average redundancy of 5.44 (Fig. S7). Crucially, all biochemical and structural assays were performed strictly on freshly purified protein; freeze-thaw cycles were found to introduce significant artifacts, particularly disrupting peptide assignments within the CTT region. Furthermore, freshly prepared ErCry4a remained entirely monomeric in solution (including under photo-excitation; Fig. S3), directly contrasting with the oligomerization previously reported in studies utilizing cryo-stored samples [44].

Upon exposure to blue light (450 nm LED, 180 μmol/m$^2$/s), ErCry4a underwent progressive photoreduction [45]. Absorption spectra tracked the sequential transition of the FAD chromophore: brief illuminations (100–600 ms) selectively populated the neutral semiquinone radical (FADH$^•$), evidenced by increased absorbance between 550 and 650 nm, while prolonged illumination (up to 10 s) drove the accumulation of the fully reduced hydroquinone (FADH$^−$), consistent with the established ErCry4a photocycle [8] (Fig. 1(D), Fig S5). Dark-state reoxidation kinetics were evaluated to determine the practical temporal window for HDX-MS labelling (Fig S4). Under aerobic conditions containing 2 mM DTT, the fully reduced state returned to the oxidized baseline with a half-life of 110 minutes. In contrast, the semiquinone state, enriched via a 600 ms illumination pulse, exhibited faster decay kinetics, with a half-life of only 28 minutes under otherwise identical conditions (Fig S6). Despite this necessarily transient nature, these lifetimes provide a sufficient temporal window to capture state-specific structural dynamics via HDX-MS. However, for the semiquinone, the maximal feasible labelling time was limited to 5 minutes. Crucially, the primary objective of this comparative HDX-MS approach is to map regions of functionally relevant structural divergence between redox states, rather than to isolate a completely homogeneous kinetic intermediate. Consequently, absolute quantitative enrichment of the semiquinone population is not strictly required; provided the partial enrichment induces statistically significant regional differences in deuterium uptake, the conformational hallmarks of the signalling state can be robustly and unambiguously defined.

### *The "fully reduced" state vs. the resting state*

We mapped the structural dynamics of the fully reduced and fully oxidized (resting) states using comparative HDX-MS. The resting state was maintained in strict darkness in HEPES buffer containing 10% glycerol and 2 mM DTT. To generate the fully reduced population, this sample was exposed to 10 s of blue light pre-illumination, and continuous ambient illumination was maintained during the labelling phase to preserve the reduced steady state. UV/Vis absorption spectroscopy indicated that this photo-activated sample initially comprised ~80% FADH$^−$ and ~20% FADH$^•$, with no detectable oxidized FAD (Fig. S5). While the statistical power for detecting subtle differences could theoretically be increased by harsher photo-reduction and the addition of chemical reducing agents, we deliberately avoided such measures to prevent photo-degradation and non-physiological structural perturbations. Given our objective to map relative regional differences rather than an absolute quantification, the realized enrichment is highly sufficient to extract robust, redox-dependent structural signatures.

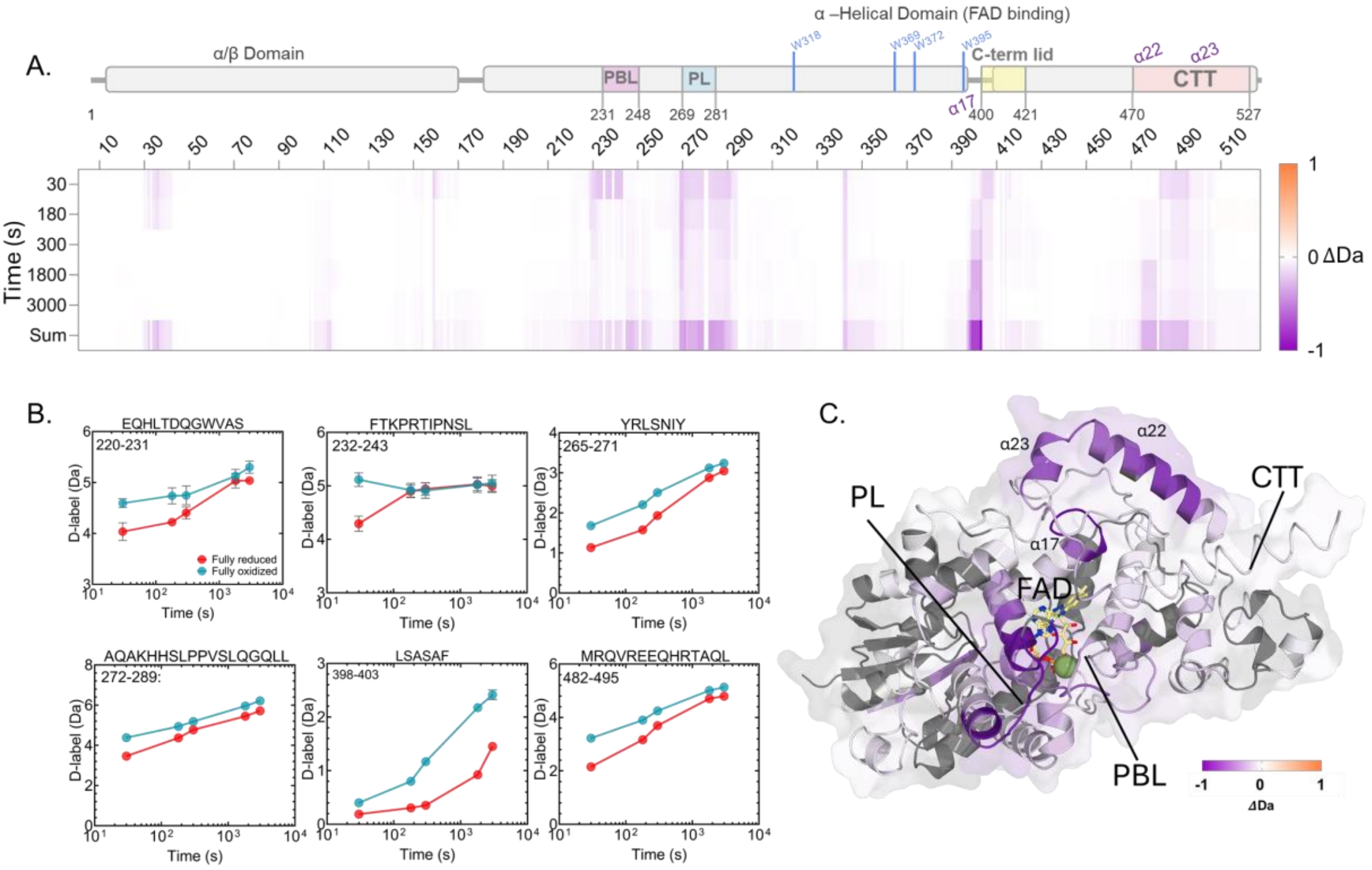

**Figure 2. HDX-MS reveals localized structural stabilization in the fully reduced state of ErCry4a.** (A) Domain organization and residue-level HDX-MS heatmap of ErCry4a. The heatmap displays the normalised differential deuterium uptake (ΔHDX) between the fully oxidized resting state and the extensively photo-reduced state across the five labelling time points. Regions exhibiting decreased exchange (protection) upon photoreduction are coloured purple (negative ΔHDX), whereas regions with increased exchange (deprotection) would be coloured orange (positive ΔHDX). Only statistically significant differences, determined via a hybrid significance testing framework [46], are plotted ($\alpha = 0.05$). (B) Deuterium uptake kinetics for representative peptides within key regions of interest. Data for the photo-reduced state and the oxidized dark state are shown in red and blue, respectively. Data points represent the mean ± 1 s.d. of three technical replicates (*n* = 5 for the 30 s time point). (C) Cumulative significant HDX differences mapped onto the ErCry4a structural model (AlphaFold2 model; FAD docked based on ClCry4a, PDB: 6PU0). Key structural nodes, including the protrusion loop (PL), phosphate-binding loop (PBL), and elements of the C-terminal tail (CTT), display enhanced protection in the photo-reduced state, consistent with light-induced conformational rigidification.

Comparative analysis of the oxidized and "fully reduced" states revealed highly specific, localized differences in deuterium uptake, indicative of major rearrangements in the solvent-accessible H-bonding network. Figure 2A maps these averaged, residue-level differences as a heatmap, calculated by subtracting the photo-reduced uptake from the dark-state uptake at each time point (see also Fig. S8 in the Supporting Material). Regions exhibiting decreased exchange (protection) upon photoactivation are coloured purple (negative ΔHDX), whereas regions with increased exchange (deprotection) are coloured orange (positive ΔHDX). Applying a hybrid significance testing framework [46], we observed exclusively protective conformational changes upon illumination; no statistically significant deprotection was detected in any peptide contig present in the mass spectra. Figure 2B illustrates the deuterium uptake time-course for representative peptides, demonstrating localized stabilization at sites of both high (e.g. α22-α23) and low (e.g. PBL) stability. Thus, the significant fully-reduced state stabilization that we observed in these highly localized parts of cryptochrome are evidenced from the primary peptide-level data.

Notably, the protrusion loop (residues 269–281) exhibited markedly reduced deuterium uptake in the fully reduced state across three overlapping peptides (Fig. 2A and 2C). This protection persisted throughout the entire labelling time course, suggesting a robust, redox-dependent stabilization of both rigidly H-bonded and transiently structured segments within this dynamic loop. However, the most profound redox-dependent difference localized to helix α17 (residues 388–399), which flanks the FAD binding pocket. Here, the photo-reduced state demonstrated strong, time-dependent protection, indicating a substantial rigidification of this helix (Fig. 2C). Given that α17 harbours the primary electron-donor tryptophan (W395) and physically bridges the FAD pocket to helix α22, its stabilization likely serves as the central allosteric hub for transmitting redox-induced structural rearrangements. Homologous FAD-proximal regions have exhibited similar HDX perturbations in non-vertebrate cryptochromes, such as CryP and CraCry [42, 43].

HDX-MS also elucidated dynamics within the PBL and the C-terminal region, elements previously implicated in ClCry4a activation [12]. Helices α22 and α23 (residues 474–494), which connect the core domain to the CTT, exhibited pronounced protection in the photo-reduced state. This stabilization localized initially to the α22–α23 loop and propagated across the entire α22 helix at longer labelling times, consistent with the tightening of an α-helical H-bonding network (Fig. 2C). Conversely, the PBL (residues 220–245) displayed distinct protection only at the earliest labelling timepoint (30 s), which rapidly attenuated. This transient protection profile is characteristic of a localized disorder-to-order transition within a highly solvent-exposed surface loop, rather than a deep structural rigidification. The distal C-terminal tail (residues 495–527) underwent near-instantaneous deuteration, confirming it remains a highly disordered and solvent-exposed domain. A marginal reduction in uptake was observed at residues 508–516, hinting at transient secondary structure or steric shielding, though the effect was modest (Fig. S8). Resolving the precise conformational kinetics of the highly dynamic PBL and CTT will ultimately require sub-second HDX methodologies.

Taken together, the data define a fully reduced state characterized by three primary nodes of persistent stabilization: the protrusion loop, the α22–α23 structural relay, and the FAD-adjacent α17 helix. These regions are either directly coupled to the chromophore or located at structural interfaces critical for conformational signalling. Meanwhile, the PBL exhibits protection, consistent with a disorder to order transition, while the bulk of the CTT remains intrinsically disordered and highly dynamic on the current experimental timescale.

### The semiquinone state exhibits a distinct, non-monotonic structural signature

Having mapped the structural endpoints of the photocycle, we next sought to determine whether the transient semiquinone ($FADH^{•}$), the presumed signalling species, induces a unique conformational state. To capture this short-lived intermediate, the resting state was photoexcited with a precise 600 ms light pulse and immediately subjected to HDX labelling in the dark. Optical absorption profiling confirmed that this protocol yielded a highly enriched population comprising ~80% $FADH^{•}$, 15% fully reduced $FADH^{-}$, and only 5% oxidized FAD (Fig S5). HDX measurements were successfully extracted at a 30 s labelling time. Notably, parallel labelling at 300 s yielded deuterium uptake patterns indistinguishable from the dark state.

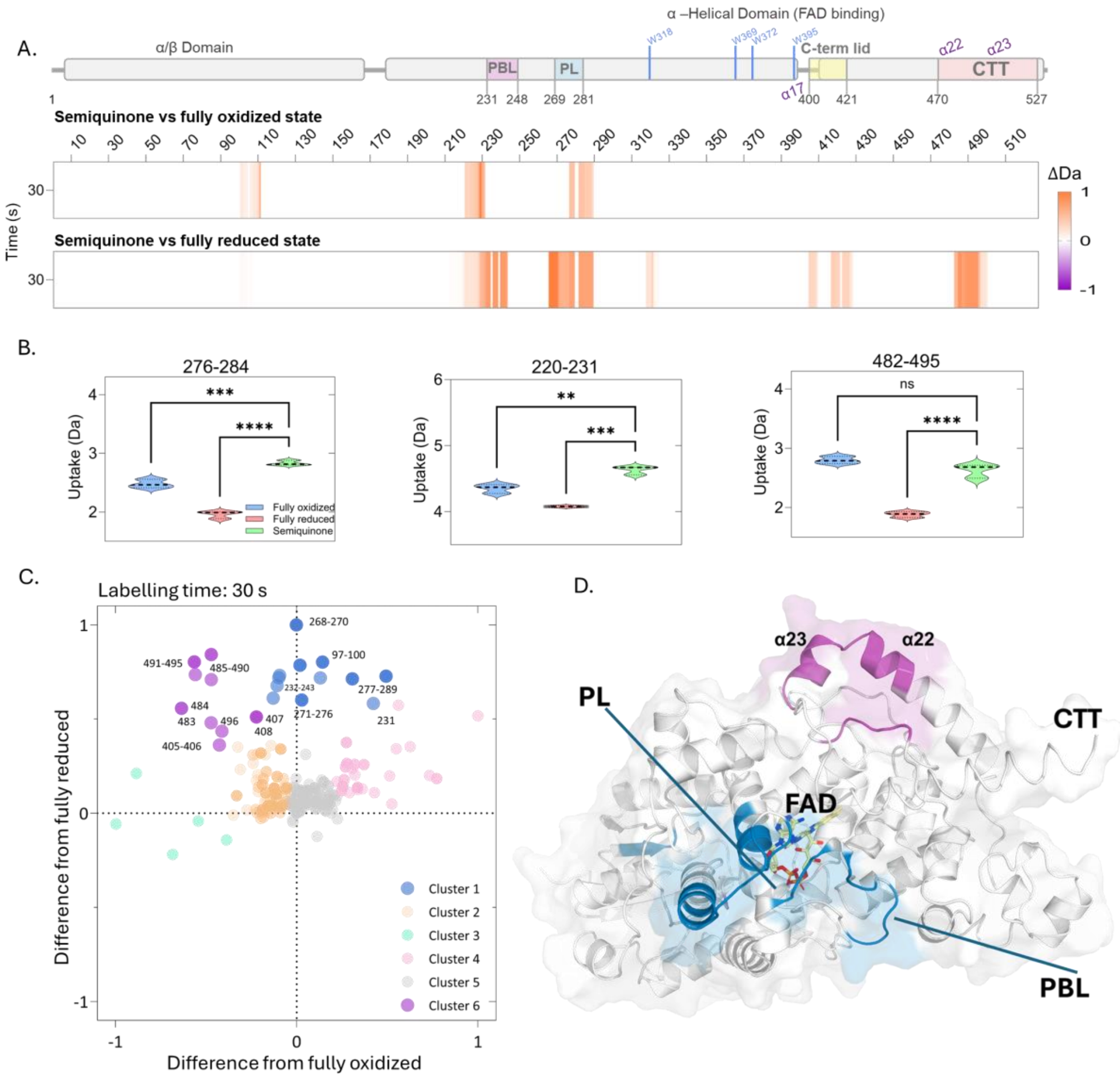

**Figure 3. The semiquinone intermediate of ErCry4a exhibits a distinct, non-monotonic structural signature.** (A) Schematic domain organization of ErCry4a aligned with residue-level HDX-MS heatmaps. The heatmaps display the differential deuterium uptake (ΔHDX) at 30 s of labelling for the semiquinone state relative to the fully oxidized dark state (SQ minus Dark) and the fully reduced state (SQ minus Fully Reduced). Regions exhibiting decreased exchange (protection) in the semiquinone state are coloured purple (negative ΔHDX), whereas regions with increased exchange (deprotection) are coloured orange (positive ΔHDX). Only statistically significant differences are shown (α = 0.05). (B) Violin plots detailing the deuterium uptake distributions for representative peptides from the phosphate-binding loop (PBL), protrusion loop (PL), and C-terminal tail (CTT) across the three defined redox states: fully oxidized (blue), semiquinone (green), and fully reduced (red). Statistical significance between the semiquinone and dark states was assessed using a two-sided Student's t-test (p-values indicated). Data represent three technical replicates. (C) 2D conformational clustering of ErCry4a structural dynamics. The sum of residue-level deuterium uptake differences is plotted comparing the semiquinone versus fully oxidized dark state (x-axis) against the semiquinone versus fully reduced state (y-axis). Residues were partitioned into six functional groups using k-means clustering. Prior to clustering, raw HDX data were centroided, corrected for the maximum number of exchangeable amide hydrogens, normalized, and pairwise state differences were calculated. (D) Structural mapping of the semiquinone-defining clusters (clusters 1 and 3) onto the predicted ErCry4a structure. The highlighted regions showcase the unique conformational footprint of the semiquinone state, with notable secondary structural elements labelled for reference.

This 30 s kinetic snapshot revealed marked, state-specific differences in deuterium uptake that unambiguously distinguish the semiquinone from both the dark and fully reduced states (Fig. 3, Fig S9-S10). Most strikingly, violin plots of peptide uptake distributions (Fig. 3B) demonstrate that the semiquinone state exhibits *increased* deuterium uptake (deprotection) in both the PBL and the PL relative to the fully oxidized dark state. This initial, transient increase in flexibility contrasts sharply with the robust protection (rigidification) these same regions undergo upon complete photoreduction to the $FADH^-$ state. Meanwhile, the early CTT, including the α22–α23 region, showed minimal perturbation in the semiquinone state, with only a slight protection observed at the extreme C-terminus (Fig. 3A). This biphasic structural trajectory, transient loosening of the PBL and PL, followed by global tightening upon full reduction, provides a much higher-resolution mechanistic picture than previous steady-state proteolysis studies CTT [12, 17], which were limited to detecting the ultimate rigidification of the PBL and CTT. Crucially, the unique deprotection signature of the semiquinone definitively proves that it is a distinct, structurally active intermediate, rather than a linear conformational blend of the dark and fully reduced states.

To broadly define the global structural signature of this intermediate, we plotted the residue-resolved deuterium uptake differences of the semiquinone versus the fully reduced state (y-axis) against the semiquinone versus the dark state (x-axis) (Fig. 3C). In this 2D conformational space, residues clustering near the origin exhibit little to no redox-state dependence. Points dispersed along the axes reflect structural changes unique to a single transition, while points located in the distal quadrants represent residues that structurally differentiate all three redox states. Using k-means clustering (Fig S11), this dynamic landscape was partitioned into six distinct structural cohorts. Cluster 1, the right-hand phase of cluster 3, and the upper region of cluster 6 most exclusively capture the unique structural footprint of the semiquinone state. When mapped back onto the ErCry4a tertiary structure (Fig. 3D), these semiquinone-defining residues heavily populate recognized functional nodes, including the PBL and PL (cluster 3), as well as the early C-terminal tail, specifically α22, α23, and their adjacent connecting loop (cluster 1). Together, these data establish the semiquinone as a structurally distinct signalling state, primed to initiate magnetoreceptive signal transduction.

Finally, we note that on the tens-of-seconds to minutes timescale required for HDX-MS labelling, any transient amino acid radicals (such as oxidized tryptophans or tyrosines) generated during the initial photoreduction have long since decayed. Consequently, the conformational dynamics mapped here are entirely attributable to the preserved FAD redox state, independent of the short-lived partner radical.

## Discussion

We have identified a series of redox-dependent structural changes in ErCry4a, revealing both well-established sites of response, such as the PBL, and additional regions not previously implicated in avian cryptochrome activation, including the α17 and α22/α23 helices. These findings demonstrate that changes in FAD redox state alone drive widespread, allosteric conformational transitions across multiple structural elements, offering a defined mechanistic route for transmitting localized magnetic information into a global biochemical signal. Notably, the semiquinone state - which forms the core of most proposed cryptochrome-based magnetoreception models, including both the photo-reduction [4] and re-oxidation hypotheses [12, 47] - emerges here as a structurally competent and functionally distinct biological entity in its own right.

Crucially, our results reveal that the semiquinone is not merely a transient, linear intermediate bridging the dark and fully reduced states; instead, it occupies a unique, non-monotonic conformational space. The regions that most clearly define its structural footprint, particularly the PBL and PL, exhibit transient destabilization and enhanced solvent accessibility. This suggests that the initial formation of the semiquinone locally loosens critical structural nodes, potentially priming the protein for downstream interaction with binding partners in the retinal cone cell. This unique structural positioning may be the very essence of its signalling capability: by encoding conformational features exclusive to one specific redox state, the semiquinone provides a mechanism for highly selective, state-dependent activation. Such selectivity is a prerequisite for a biological magnetoreceptor, enabling precise, light-driven signal gating in a protein system otherwise capable of continuous redox cycling.

These state-resolved findings provide necessary nuance to earlier models of cryptochrome signalling. Previous limited proteolysis studies on pigeon ClCry4 [12] were interpreted as evidence for a uniform, light-induced protection of both the PBL and the CTT. In contrast, our temporally resolved HDX-MS data reveal that such widespread protection (which we interpret as thermodynamic stabilization) occurs exclusively in the fully reduced state. The semiquinone intermediate, conversely, exhibits distinct *deprotection* of the PBL, while the CTT remains highly dynamic and solvent-accessible, save for its most proximal segment. We propose that these apparent discrepancies are rooted in methodological resolution rather than fundamental biological differences. Steady-state limited proteolysis lacks the kinetic resolution required to trap a short-lived transient intermediate; consequently, earlier experiments likely captured the time-averaged accumulation of the long-lived, fully reduced state, inadvertently bypassing the structural realities of the semiquinone. By directly deconvoluting these redox-dependent structural fingerprints under strictly controlled conditions, HDX-MS resolves this ambiguity, establishing the transient local flexibility of the semiquinone as the likely trigger for magnetoreceptive signalling.

The observed transient increase in PBL flexibility qualitatively aligns with previous molecular dynamics (MD) simulations conducted at both the all-atom and coarse-grained levels [48-51]. Schuhmann et al. reported that the PBL opens upon formation of the initial radical pair ($FAD^{\bullet-}$ and $Trp^{\bullet+}$) [49]. Subsequent coarse-grained simulations suggested a "reclosure" of this loop occurring on the 100 µs timescale [50]. Given the inherent temporal limitations of MD in capturing slower, large-scale structural dynamics, this computationally observed "reclosure" may reflect localized conformational sampling rather than a true thermodynamic stabilization. With this caveat, the MD trajectories strongly complement our HDX-MS data, which directly demonstrate enhanced PBL motility in response to the redox-driven charge shifts of semiquinone formation. Crucially, our empirical measurements reveal that these fast PBL dynamics are allosterically coupled to the adjacent protrusion loop (~10 Å away), which undergoes comparable transient destabilization. The ultimate rigidification of these regions, however, is exclusively captured in the fully reduced state and on a much slower timescale (cf. Fig. 2). This temporal disparity could explain why such long-range coupling and ultimate stabilization were not detected in previous computational models constrained by shorter simulation windows.

A particularly striking structural divergence occurs at helix α22, located near the terminal tryptophan (Trp369) of the electron-transfer cascade. Previous HDX-MS investigations of the algal cryptochromes CryP [42] and CraCry [43] highlighted this same region; however, in those species, photoactivation induced *deprotection* (destabilization) rather than the pronounced *protection* we observe in ErCry4a. For CraCry, this destabilization was attributed to a disrupted interaction network between helix α22 and the adjacent protein core, specifically, electron-transfer-dependent protonation events (involving Tyr373 and Asp321) disrupt specific salt bridges, causing the helix to unfold [18]. We propose that this species-specific divergence in ErCry4a is rooted in its differing local electrostatics during the transient formation of the flavin-tryptophan radical pair. In both CryP and CraCry, helix α22 carries a net positive charge (+4e), whereas the α22 helix of ErCry4a yields a net charge of –1e. Because ErCry4a utilizes a different terminal radical ($TrpH^{\bullet+}$) and features this inverted local charge profile, it likely avoids the specific salt-bridge disruption seen in CraCry, instead allowing local electrostatics to drive the swift functional rigidification of the avian magnetoreceptor. While attributing complex HDX shifts solely to electrostatic potentials is a simplification and the timescales of radical pair formation and the structural transitions identified here are widely different, this fundamental inversion of the local charge environment and a swift local response to it provides a plausible structural rationale for why the avian magnetoreceptor exhibits rigidification (protection) of α22 rather than the destabilization seen in its algal homologs.

Finally, we must consider whether the protection patterns observed in the fully reduced state arise from protein dimerization. Oligomerization is a well-documented mechanism of cryptochrome photoactivation, most notably in plant cryptochromes [27, 28], and recent computational work has proposed several putative dimerization interfaces for ErCry4a [44]. However, multiple lines of experimental evidence argue against oligomerization driving our observed structural changes. First, none of the protected regions identified in our HDX-MS dataset directly map to these predicted dimer interfaces. Second, the pronounced flexibility of both the PL and PBL actively precludes their participation in a stable dimer interface during the semiquinone state. Third, and most definitively, native size-exclusion chromatography confirmed that our freshly prepared, fully photo-reduced samples remain strictly monomeric in solution. Furthermore, the extreme flexibility of the distal CTT makes it highly unlikely that the observed protection of the PBL arises from an intramolecular back-folding of the tail to occlude this site. We therefore conclude that the structural transitions mapped here represent the intrinsic intramolecular allostery of the monomeric protein. While previous studies utilizing cryo-stored or aged samples have reported marked oligomerization [44], our data firmly establish that the primary, redox-driven structural signalling event in fresh ErCry4a is fundamentally a monomeric process. If higher-order oligomerization is indeed functionally relevant *in vivo*, it likely represents a secondary, downstream consequence of these initial monomeric conformational changes.

Finally, when considering the ultimate output of the cryptochrome photocycle, it is important to acknowledge that *in vivo* signal transduction need not be exclusively structural. Photochemical redox cycling could also propagate signals indirectly via the localized generation of reactive oxygen species (ROS), which are intrinsic byproducts of flavin dark-state re-oxidation. This alternative paradigm has recently gained traction in *D. melanogaster* models, where an isolated C-terminal fragment, entirely lacking the canonical FAD-binding pocket, was shown to be sufficient to mediate magnetic field effects [52]. However, whether these ROS-driven effects represent a primary, high-fidelity magnetic sensor or a secondary chemical consequence of ambient light exposure remains a subject of debate. By contrast, our mapping of robust, state-specific structural dynamics in ErCry4a provides unambiguous biophysical evidence for a direct, allosteric mode of signal transduction. By demonstrating that the transient semiquinone state possesses a unique, conformationally active structural signature, our data establish a definitive mechanistic link between quantum-level spin dynamics and global protein topology. This strongly supports the existence of a dedicated, protein-mediated signalling cascade, i.e. a high-fidelity signal transduction pathway, that perfectly aligns with the rigorous physiological demands of the evolutionarily optimized compass sense in night-migratory songbirds.

## Conclusions

Our state-resolved HDX-MS analyses demonstrate that ErCry4a undergoes distinct, redox-dependent structural transitions, defined by key allosteric shifts in the phosphate-binding loop (PBL), protrusion loop (PL), the FAD-proximal α17 helix, and the α22/α23 network. These results validate previously implicated structural motifs while uncovering entirely new dynamic elements, underscoring that localized photochemical redox transitions are strictly coupled to the global conformational responses required to transmit magnetic information.

Crucially, the semiquinone state emerges not as a simple, linear structural stepping-stone between the resting and fully reduced states, but as a functionally competent and conformationally unique biological entity. It is characterized by an initial, transient destabilization of the PBL and PL, contrasting sharply with the robust rigidification observed in the fully reduced form. This non-monotonic structural trajectory establishes the semiquinone as a uniquely active signalling state, consistent with its central role in leading biophysical models of cryptochrome magnetoreception. Furthermore, our data refine earlier structural interpretations by demonstrating that CTT compaction is strictly localized to its most proximal segment, while the distal tail remains intrinsically disordered and highly dynamic, providing a high-resolution update to previous structural models.

Together, these findings provide direct biophysical evidence that the ErCry4a photocycle drives the robust, allosteric structural dynamics necessary to support a high-fidelity protein signalling pathway. While secondary ROS-mediated metabolic effects cannot be entirely ruled out *in vivo*, the highly specific, state-dependent structural choreography we observe strongly favours a dedicated, protein-mediated mechanism of sensory transduction. These results firmly establish ErCry4a as a structurally responsive magnetosensor, advancing our fundamental understanding of how quantum-level photochemistry is harnessed for navigation in migratory birds. Future studies will employ sub-second temporally resolved HDX-MS to capture the ultra-fast kinetics of these highly mobile domains and extend this powerful comparative approach to other putative magnetoreceptor candidates, such as ErCry1a, to further map the evolutionary landscape of this remarkable sensory modality.

## Methods

### Chemicals and Reagents

Chemicals were purchased as follows: Tris(hydroxymethyl)aminomethane (Tris), sodium chloride (NaCl), potassium phosphate monobasic and dibasic, and urea were purchased from Sigma-Aldrich. Deuterium oxide ($D_2O$, 99.9 atom % D) was obtained from Cambridge Isotope Laboratories. Leucine enkephalin (used as a lock mass reference) was purchased from Waters. Water, acetonitrile, and formic acid (99.5%) Optima LC/MS Grade were obtained from Fisher Scientific. All other ultrapure water used was purified on a Milli-Q Advantage A10 system (Merck).

### Cloning, expression and purification of ErCry4a

The purification of ErCry4a was carried out as described in ref. [8] with slight modification. Briefly, the full-length coding sequence of ErCry4a (GenBank: KX890129.1) was synthesised and cloned (Genscript) into the pCold I expression vector (Takara Bio) between the NdeI and XhoI restriction sites for expression in *E. coli* SoluBL21 cells (Genlantis). Overexpression was achieved by growing cells in TB medium at 24°C to an $A_{600}$ of 0.6, followed by induction with 5 μM IPTG for 20 h. For purification, the cells were sedimented and lysed with lysis buffer [50 mM Tris (pH 8.0), 500 mM NaCl, 20% glycerol (vol/vol), 10 mM imidazole and 2 mM DTT, and one EDTA-free protease inhibitor tablet (Roche) added per 50 mL of buffer]. The lysed cells were sonicated and centrifuged for 45 min at 16,743 ×g at 4°C. The supernatant was mixed with pre-equilibrated Ni-NTA resin for 1.5 h at 4°C with gentle rotation followed by gravity-flow column chromatography. The eluted fractions were pooled, concentrated and diluted to 50 mM NaCl. The protein was then passed over a 1 mL Hitrap Q HP anion exchange column which had been equilibrated with Start buffer [50 mM HEPES (pH 8.0), 50 mM NaCl, 10 % glycerol (vol/vol) and 2 mM DTT]. ErCry4a was eluted using 7.5-50% gradient of elution buffer [50 mM HEPES (pH 8.0), 1 M NaCl, 10 % glycerol (vol/vol) and 2 mM DTT] over 40 CV at 0.5 mL/min. The elution peak was observed at the conductivity of ca. 12 mS/cm and at 180 mM NaCl concentration. Fractions were concentrated and used for assays. All protein preparation and purification steps were carried out in the dark or under dim red light to minimise unintended light exposure. During sample handling and storage, proteins were kept in dark amber coloured Eppendorf tubes or Falcon tubes to protect them from ambient light until the intended illumination during the experiments.

### Dimerization test

An analytical column (Superose 6 Increase 10/300 GL) was used to analyse the formation of dimers upon light activation of ErCry4a. The buffer containing 50 mM HEPES, 180 mM NaCl, 10% glycerol, and 2 mM DTT was used for column equilibration. Samples were injected using a 500 µl loop and run at a flow rate of 0.5 mL/min. For analysis of the oxidised (dark) state, the column was protected from light by wrapping it in foil. For the photoreduced state, the sample was illuminated with 450 nm light at 3.41 $mW/cm^2$ for 10 s using a blue light emitting diode (Thorlabs M450LP2) prior to loading onto the column.

### UV/vis absorption spectroscopy

The UV-visible experiments were recorded on an Agilent Cary 60 UV spectrophotometer at room temperature in the 300 - 800 nm range. All spectra were blanked against buffer containing 50 mM HEPES, 180 mM NaCl, 10% glycerol and 2 mM DTT at pH 8. The light source for photoreduction consisted of a blue light emitting diode with emission wavelength of 450 nm (Thorlabs M450LP2), driven with a 800 mA current (40%) to produce an irradiance of 3.41 $mW/cm^2$. Light power was measured using a photodiode sensor positioned at the cuvette position (Thorlabs, PM400 & S120C). The absorption spectra were recorded in the dark and light pulses of various length were used to determine the illumination required for ErCry4a to reach the semiquinone state under the experimental conditions.

**Reoxidation kinetics by UV/vis absorption spectroscopy**

To monitor reoxidation kinetics, the protein sample (50 mM HEPES, 180 mM NaCl, 10% glycerol, 2 mM DTT, pH 8) was photo-reduced using illumination at 3.41 $mW/cm^2$ for 600 ms, generating the maximum population of the semiquinone state; for the fully reduced state an illumination of 10 s was used. UV-visible spectra were then recorded at defined time intervals in the dark to follow the return to the oxidized state. The half-life of the semiquinone was obtained by fitting the decay kinetics to a monoexponential decay model in GraphPad Prism.

**Hydrogen–Deuterium Exchange**

All labelling experiments were performed at 23°C, further quenched, and analysed at 0°C. Protein samples (10 µM) were labelled at five time points (30 s, 180 s, 300 s, 30 min and 50 min), $n$ = 3. Buffers used were 50 mM Tris, 150 mM NaCl at pH 8.00 in $H_2O$; 50 mM Tris, 150 mM NaCl at pD 8.00 (pH meter reading 7.59) in $D_2O$, and 100 mM potassium phosphate, 3 M urea, pH 2.54 in $H_2O$, as equilibrium, labelling and quench buffers, respectively. For each timepoint, three technical replicates were performed, with measurements taken on the same day (or overnight) to ensure consistency. For the semiquinone state, a fresh sample was used for each timepoint; for the fully reduced state the sample was renewed after every 4 hours and labelling times were assigned randomly. Labelling was initiated by a 20-fold dilution of the protein sample into 99.9% $D_2O$ labelling buffer, yielding a final $D_2O$ content of ~95%. Sample mixing, labelling, and quenching were performed using a LEAP HDX PAL robotic system (LEAP Technologies, now Trajan Scientific), coupled directly to the Waters HDX Manager, to ensure reproducible handling and minimize back-exchange.

**Liquid Chromatography–Mass Spectrometry**

Samples were digested online for 3 min at 90 µL/min with a Nepenthesin-2/Pepsin-immobilized pepsin column (AffiPro), then trapped on a VanGuard 2.1 × 5 mm ACQUITY BEH C18 column (Waters) and separated on a 1 × 100 mm ACQUITY BEH 1.7 µm C18 column (Waters) with a 7 min linear gradient of acetonitrile (5-40%) supplemented with 0.1% formic acid. Peptides were eluted into a Synapt G2-Si mass spectrometer (Waters). Mass spectra were obtained using the Waters $HDMS^E$ mode within a mass range of 150 to 2000 m/z. Instrument parameters were configured as follows: a capillary voltage of 3.0 kV, a cone voltage of 40 V, a trap collision energy of 6 V, a travelling wave ion mobility separation at a velocity of 474 m/s, a wave amplitude of 36.5 V, and a nitrogen pressure of 2.75 mbar. For low-energy scans, a transfer collision energy of 6 V was applied, while high-energy scans utilized three separate collision energy ramps ranging from 25 to 55 V.

**Data Analysis**

ProteinLynx Global Server 3.03 (Waters) was used to identify peptides discoverable during analysis. Deuterium incorporation was determined with DynamX 3.0 (Waters), with a manual review of all assignments. Other statistical tests, difference and clustering analysis were performed with the in-house software tool HydroBot [53]. Structures were modelled in PyMOL (Schrödinger).

**Acknowledgements**

This work was supported by the Biotechnology and Biological Sciences Research Council (grant numbers BB/Y514147/1 and 2717616). We thank Drs. Archipowa and Kutta for hosting S.P. during a visit at University of Regensburg and their comments on an early draft of this manuscript. For the purpose of open access, the author has applied a Creative Commons Attribution (CC BY) licence to any Author Accepted Manuscript version arising from this submission.